\documentclass[10pt,preprint]{aastex}

\def\be{\begin{eqnarray}}
\def\ee{\end{eqnarray}}

\def\bh{_{\rm BH}}
\def\bol{_{\rm bol}}
\def\Edd{_{\rm Edd}}
\def\kj{^{j,k}}
\def\Sp{_{\rm Sp}}
\def\mod{_{\rm mod}}
\def\obs{_{\rm obs}}
\def\OIII{_{\rm [OIII]}}

\def\dex{{\rm\,dex}}
\def\ergs{{\rm\,erg\,s^{-1}}}
\def\kms{{\rm\,km\,s^{-1}}}
\def\Lsun{{\rm\,L_\odot}}
\def\msun{{\rm\,M_\odot}}

\def\kpc{{\rm\,kpc}}
\def\yr{{\rm\,yr}}
\def\pc{{\rm\,pc}}

\begin{document} \title{Evolution of accretion disks around massive black
holes: constraints from the demography of active galactic nuclei} 
\author{Qingjuan Yu\footnotemark[1] $^{,2}$, Youjun Lu$^{2}$ and Guinevere
Kauffmann$^3$} \affil{$^2$Department of Astronomy and Theoretical Astrophysics
Center, University of California at Berkeley, Berkeley, CA 94720; yqj@astro.berkeley.edu; lyj@astro.berkeley.edu}
\affil{$^3$Max-Planck-Institut f$\ddot{u}$r Astrophysik, D-85748 Garching,
Germany; gamk@mpa-garching.mpg.de
} \footnotetext[1]{Hubble Fellow} \begin{abstract}

Observations have shown that the Eddington ratios (the ratio of the bolometric
luminosity to the Eddington luminosity) in QSOs/active galactic nuclei (AGNs)
cover a wide range.  In this paper we connect the demography of AGNs obtained
by the Sloan Digital Sky Survey with the accretion physics around massive black
holes and propose that the diversity in the Eddington ratios is a natural
result of the long-term evolution of accretion disks in AGNs.  The observed
accretion rate distribution of AGNs (with host galaxy velocity dispersion
$\sigma\simeq70-200\kms$) in the nearby universe ($z<0.3$) is consistent with
the predictions of simple theoretical models in which the accretion rates
evolve in a self-similar way.  We also discuss the implications of the results
for the issues related to self-gravitating disks, coevolution of galaxies and
QSOs/AGNs, and the unification picture of AGNs.

\end{abstract} \keywords{black hole physics --- accretion, accretion disks ---
galaxies: active --- galaxies: evolution} \maketitle

\section{Introduction}\label{sec:intro}

The past decade has seen great progress on measuring the mass of massive black
holes (BHs) in both nearby normal galaxies and AGNs/QSOs (e.g., \citealt{M98,R98};
\citealt{W99}; \citealt{K00}). With reliably measured BH masses in AGNs/QSOs,
their Eddington ratios (i.e., the ratio of the bolometric luminosity to the
Eddington luminosity) can be estimated. Compared to the bolometric luminosity
(or the accretion rate, given the mass-to-energy conversion efficiency
$\epsilon$), the Eddington ratio (or dimensionless accretion rate) is a
normalized parameter and more related to the accretion physics.  Observations
have shown that the Eddington ratios in AGNs/QSOs cover a wide range (at least
2--3 orders of magnitude) with an upper limit of about 1 (\citealt{WU02};
\citealt{MD04}). It is natural to ask what determines the diversity of the
Eddington ratios. The simplest attempt to answer this question would be by
first asking whether the diversity of the Eddington ratios is a consequence of
evolution of accretion processes onto BHs.

The questions raised above on the origin of the diversity of the Eddington
ratios in AGNs/QSOs may have important implications for understanding the
evolution of the AGN/QSO luminosity function (e.g., \citealt{SG83};
\citealt{B00}).  Generally the number density of AGNs/QSOs at a given redshift
and luminosity/dimensionless accretion rate should be jointly determined by
both the number densities of galaxies whose nuclear activities are triggered at
earlier time and the follow-up nuclear luminosity/accretion rate evolution of
individual AGNs/QSOs. The history of the triggering rate is related to the
formation and evolution of galaxies (e.g., major mergers of galaxies) and may
have a strong cosmic evolution (e.g., \citealt{KH00}). The follow-up nuclear
luminosity/accretion rate evolution of individual AGNs/QSOs is directly related
to the evolution of the accretion disk around the BH and the fueling process
onto the disk in their host galaxies. In this paper we extract the evolution of
the accretion processes from the observational distribution of accretion rates
in AGNs. This provides constraints on theoretical models of accretion disk
evolution and also illuminates the origin of the diversity of the Eddington
ratios.

To extract the information on the nuclear luminosity/accretion rate evolution
from observations, statistical methods  involving a large sample of AGNs are
required since (i) a single AGN may only represent one specific period in a
prolonged phase of nuclear activity.  A large sample of AGNs with different
ages will span all phases of this activity  and allow us to extract information
about evolution; and (ii) in addition to age, other physical parameters may be
important in determining how AGNs evolve and a statistical method may help to
clarify these.  In \S~\ref{sec:methods}, we describe our method of extracting
the evolution of accretion processes phenomenologically from the observational
distribution of the accretion rates in nearby AGNs (see also the related
methods in \citealt{YL04a,YL04b}). We argue that results for nearby AGNs are
not strongly affected by the uncertainty in the nuclear activity triggering
history of these objects, because  the evolution of the triggering rate at low
redshifts ($z<0.5$) is probably rather weak. Recent progress by the Sloan
Digital Sky Survey (SDSS) on the demography of AGNs
(http://www.mpa-garching.mpg.de/SDSS/; \citealt{K03b,H04}) makes the
application of the method practical. In \S~\ref{sec:obs}, we apply the method
to a large sample of SDSS AGNs and get the phenomenological constraints on the
evolution of accretion processes.  We find that the constraints are consistent
with theoretical expectations of the accretion disk evolution models shown in
\S~\ref{sec:models}. This consistency suggests that the diversity of Eddington
ratios in AGNs is probably a natural result of the evolution of the accretion
disks around massive BHs.  In \S~\ref{sec:dis}, we also discuss the implications
of the results for the long-term evolution of accretion disks, AGN unification
models, and coevolution models of galaxies and QSOs/AGNs etc. Our conclusions
are summarized in \S~\ref{sec:con}.

\section{Methodology}\label{sec:methods}

In this section we analyze how the accretion rate evolution of an AGN is
incorporated into the accretion rate distribution of AGNs in the local
universe.

Observations have revealed that the central BH mass in nearby normal galaxies
is tightly correlated with the velocity dispersion of the bulge components of
host galaxies \citep{FM00,G00,T02}. In this paper we take this correlation
as a boundary condition for mass growth of BHs in AGNs, that is, the final BH
masses of nearby AGNs after the nuclear activity, $M_f$, is assumed to follow
the same correlation with
velocity dispersion $\sigma$, since mass growth of BHs comes mainly from the
nuclear activity phase \citep[e.g.,][]{YT02,F04,M04}. Here the velocity
dispersion $\sigma$ (or the gravitational potential field in galactic scales)
is assumed to not change significantly during and after the nuclear activity
phase. This assumption has also been implicitly used in many physical models to
interpret the correlation between BH mass and velocity dispersion in nearby
normal galaxies (e.g., \citealt{SR98,B99,F99,AGR01,BS01,King03,MQT05}; see
also more discussions on the BH mass versus velocity dispersion relation in AGNs
in \S~\ref{sec:dis} and in \citealt{YL04b}). We describe the
correlation through a probability distribution function of $\log M_f$ at a
given $\sigma$, $P(\log M_f|\sigma)$, and $P(\log M_f|\sigma)d\log M_f$ gives
the probability that the logarithm of the BH mass of an AGN with velocity
dispersion $\sigma$ is in the range $\log M_f\rightarrow \log M_f+d\log M_f$.
The mean $\log M_f$ at a given $\sigma$ is denoted by $\langle \log
M_f|\sigma\rangle$[$=8.13+4.02\log(\sigma/200\kms)$; \citealt{T02}].  The
intrinsic scatter of $P(\log M_f|\sigma)$ around $\langle \log
M_f|\sigma\rangle$ is small enough that it is difficult to distinguish from
measurement errors, but is less than $0.27\dex$ \citep{T02}. Given the
bolometric luminosity of an AGN $L\bol$, the mass accretion rate is given by
$\dot M\bol\equiv L\bol/(\epsilon c^2)$, where $c$ is the speed of light. In
practice, the [OIII] $\lambda 5007$ line luminosity $L\OIII$ is often used as a
tracer of AGN activity (e.g., see \citealt{H04}); and the bolometric luminosity
of an AGN may be estimated from its observed [OIII] line luminosity, given its
bolometric correction $f\bol=L\bol/L\OIII$.  We may also define the accretion
rate through the [OIII] line luminosity by $\dot M\OIII\equiv f'\bol
L\OIII/(\epsilon c^2)$, where $f'\bol$ is a factor representing the average
bolometric correction between $L\OIII$ and $L\bol$ and given through $\log
f'\bol=\langle\log f\bol\rangle$. The average of the logarithm of the
bolometric correction $\langle\log f\bol\rangle$ may be obtained by giving the
probability distribution function $P(\log L\OIII|L\bol)$, which is defined so
that $P(\log L\OIII|L\bol)dL\OIII$ is the probability that the logarithm of the
[OIII] line luminosity is in the range $\log L\OIII\rightarrow \log
L\OIII+d\log L\OIII$ at a given $L\bol$.  In general, $\dot M\bol$ and $\dot
M\OIII$ are not the same since the bolometric corrections are not the same for
all the AGNs.

Given an AGN with nuclear activity triggered at cosmic time $t_i$ and with
final BH mass $M_f$, we describe the evolution of its accretion rate by $\dot
{\cal M}\bol(\tau,t_i,M_f)$, where $\tau\equiv t-t_i$ is the physical time
passed since the triggering of the nuclear activity.  The total time that the
AGN will spend with the logarithm of its accretion rate in the range $\log\dot
M\bol\rightarrow \log\dot M\bol+d\log\dot M\bol$ at cosmic time $t_l\equiv
t(z=z_l)\le t\le t(z=0)\equiv t_0$ is given by (see eq.~15 in \citealt{YL04a}):
\begin{eqnarray}
& & T(\dot M\bol,t_i,M_f)d\log \dot M\bol \nonumber \\
&\equiv & \sum_k\frac{(\dot M\bol\ln10)d\log \dot M\bol}
{\left|d \dot {\cal M}\bol(\tau,t_i,M_f)/d \tau|_{\tau=\tau_k}\right|},
\label{eq:PLM}
\end{eqnarray}
where $\tau_k$ ($k=1,2,...$) are the solutions of the equation $\dot {\cal
M}\bol(M_f,\tau,t_i)=\dot M\bol$ that satisfy $t_l\le t_i+\tau_k\le t_0$, and
$T(\dot M\bol,t_i,M_f)=0$ if no such solutions exist.  In this paper the
redshift $z_l$ is set to 0.3 and AGNs with $z\le z_l$ are taken as nearby AGNs.
We define the function $\phi(\dot M\bol,M_f,\sigma,t)$ so that $\phi(\dot
M\bol,M_f,\sigma,t)d\log M\bol d\log M_f d\log\sigma$ gives the comoving number
density of AGNs at redshift $z(t)$ with the logarithms of the accretion rate,
the final BH mass, and the velocity dispersion in the ranges $\log\dot
M\bol\rightarrow\log\dot M\bol+d\log\dot M\bol$, $\log M_f\rightarrow\log
M_f+d\log M_f$, and $\log\sigma\rightarrow\log\sigma+d\log\sigma$.  Given the
evolution of the nuclear activity triggering rate $N(t_i,M_f,\sigma)$, which is
defined so that $N(t_i,M_f,\sigma)dt_id\log M_f d\log\sigma$ is the comoving
number density of BHs with host galaxy nuclear activity triggered at cosmic
time $t_i\rightarrow t_i+dt_i$ and with the logarithms of its final BH mass and
velocity dispersion in the ranges $\log M_f\rightarrow \log M_f+d\log M_f$ and
$\log\sigma\rightarrow\log\sigma+d\log\sigma$, we have the time integral of
$\phi(\dot M\bol,M_f,\sigma,t)$ given by (cf., eq.~10 in \citealt{YL04a})
\begin{eqnarray}
& & \int_{t_l}^{t_0}\phi(\dot M\bol,M_f,\sigma,t)dt \nonumber \\
& = & \int_0^{t_0} N(t_i,M_f,\sigma)T(\dot M\bol,t_i,M_f)d t_i\nonumber \\
& = & \mathbb{N}(\dot M\bol,M_f,\sigma)\mathbb{T}(\dot M\bol,M_f),
\label{eq:phit}
\end{eqnarray}
where 
\be
\mathbb{N}(\dot M\bol,M_f,\sigma)\equiv \frac{\int_0^{t_0} N(t_i,M_f,\sigma)T(\dot M\bol,t_i,M_f)d t_i}{\int_0^{t_0}T(\dot M\bol,t_i,M_f)d t_i},
\ee
and
\be
\mathbb{T}(\dot M\bol,M_f) & \equiv & \int_0^{t_0}T(\dot M\bol,t_i,M_f)d t_i.
\label{eq:T}
\ee
By multiplying equation (\ref{eq:phit}) by $P(\log L\OIII|L\bol)$ and then
integrating it over $\log \dot M\bol$ and $\log M_f$, we have
\begin{eqnarray}
& & \int_{t_l}^{t_0}\Phi(\dot m\OIII,\sigma,t)dt
\nonumber\\
&=& \int d\log M_f \int d\log\dot M\bol P(\log L\OIII|L\bol)
\mathbb{N}(\dot M\bol,M_f,\sigma)\mathbb{T}(\dot M\bol, M_f),
\label{eq:Phit}
\end{eqnarray}
where 
\be
\Phi(\dot m\OIII,\sigma,t)=\int d\log M_f\int d\log\dot M\bol P(\log L\OIII|L\bol)\phi(\dot M\bol,M_f,\sigma,t),
\label{eq:Phi}
\ee
$\dot m\OIII\equiv \dot
M\OIII/\dot M\Edd[M_f'(\sigma)]$ is the normalized accretion rate,
$M_f'(\sigma)=10^{\langle\log M_f|\sigma\rangle}$, $\dot
M\Edd[M_f'(\sigma)]\equiv L\Edd[M_f'(\sigma)] /(\epsilon c^2)$, and
$L\Edd[M_f'(\sigma)]$ is the Eddington luminosity of a BH with mass
$M_f'(\sigma)$.
Note that here the definition of the normalized accretion rate is a little
different from the conventionally defined Eddington ratio in that $\dot M\OIII$
may not be the same as $\dot M\bol$, and the Eddington luminosity used here is
for the {\em average final} BH mass, rather than for the BH mass that is
powering the nuclear activity in an AGN. If the bolometric correction of each
AGN in the sample and its BH mass that is powering the nuclear activity are
obtained from future observations, we may construct the methods for the
distribution of the Eddington ratios as above and perform a consistency check
on the results.

We can further define the time-averaged accretion rate and velocity
distribution function of nearby AGNs (with $z\le z_l$) by
\be
\bar{\Phi}(\dot m\OIII,\sigma)\equiv \frac{\int_{t_l}^{t_0}\Phi(\dot
m\OIII,\sigma,t)dt}{t_0-t_l}.
\label{eq:barPhi}
\ee
Below we assume that the triggering rate of nearby AGNs
$N(t_i,M_f,\sigma)=n(t_i,\sigma)P(\log M_f|\sigma)$, where
$n(t_i,\sigma)=\int d\log M_f N(t_i,M_f,\sigma)$ is independent of
$M_f$. By further assuming that $n(t_i,\sigma)$ is independent of $t_i$, we
have $\mathbb{N}(\dot M\bol,M_f,\sigma)=N(t_i,M_f,\sigma)$; and
then by combining equations (\ref{eq:Phit}) and (\ref{eq:barPhi}), we have
\begin{eqnarray}
& & \bar{\Phi}(\dot m\OIII,\sigma)\nonumber\\
&=& n(\sigma)\int d\log M_f P(\log M_f|\sigma)\int d\log\dot M\bol P(\log L\OIII|L\bol) \frac{\mathbb{T}(\dot M\bol, M_f)}{t_0-t_l}.
\label{eq:Phit1}
\end{eqnarray}
Below we also assume that $\dot {\cal M}\bol(\tau,t_i,M_f)$ is independent of
the cosmic time $t_i$, and according to equations (\ref{eq:PLM}) and
(\ref{eq:T}), we have
\be
\mathbb{T}(\dot M\bol, M_f)&=& \sum_k\int^{t_0-\tau_k}_{\min\{0,t_l-\tau_k\}}\frac{\dot M\bol\ln10} {\left|d \dot {\cal M}\bol(\tau,M_f)/d \tau|_{\tau=\tau_k}\right|}dt_i,
\ee
where $\tau_k$ ($k=1,2,...$) are the solutions of the equation $\dot {\cal
M}\bol(M_f,\tau)=\dot M\bol$ that satisfy $0\le\tau_k\le t_0$.
If all the solutions $\tau_k<t_l$ ($k=1,2,...$; which is satisfied by the assumed models and
its best-fit parameters shown in Fig.~\ref{fig:contour} below),
we have $\mathbb{T}(\dot
M\bol, M_f)\propto t_0-t_l$, $\bar{\Phi}(\dot
m\OIII,\sigma)=\Phi(\dot m\OIII,\sigma,t)$ at $z(t)\le z_l$, and $\Phi(\dot
m\OIII,\sigma,t)$ is independent of $t$.  
In reality AGNs may have a high
triggering rate at high redshifts, however, AGNs triggered at high redshifts
may have become faint enough at the current epoch to move out of the range of
the accretion rates of  interest in this paper.  For example, using the
accretion rate evolution model assumed below (eqs.~\ref{eq:Lphase1} and
\ref{eq:Lphase2}) and the best-fit parameters obtained in \S~\ref{sec:obs},
only those AGNs whose nuclear activities are triggered at $z<0.5$ may have an
accretion rate high enough at $z<z_l$ to be within the range shown in Figure
\ref{fig:distr}a below.  As an additional check, we have assumed  that the
triggering rate is proportional to the major merger rate of galaxies, e.g.,
$n(t_i,\sigma)\propto[1+z(t_i)]^\alpha$ with $\alpha=0.51\pm0.28$ \citep{L04},
and our results are not significantly affected.

It is plausible to assume that the growth of a BH involves two phases after the
nuclear activity is triggered on (\citealt{YL04a}; see also \citealt{SB92}).
In the first (or ``demand limited'') phase, there is plenty of material to fuel
BH growth. However, not all of the available material can contribute to the BH
growth immediately and the BH growth is limited by the Eddington luminosity.
We assume that the BH accretes with the Eddington luminosity for a period
$\tau_I$ and denote the BH mass at cosmic time $t_I=t_i+\tau_I$ by $M_I$.  We
assume the mass-to-energy conversion efficiency $\epsilon$ to be a constant.
Thus, the accretion rate in the first phase increases with time as follows:
\be
\dot{\cal M}\bol(\tau)=\dot M\Edd(M_I)
\exp\left(\frac{\tau-\tau_I}{\tau\Sp}\right),
\qquad 0<\tau<\tau_I,
\label{eq:Lphase1}
\ee
where $\tau\Sp=4.5\times 10^7\epsilon/[0.1(1-\epsilon)]\yr$ is the Salpeter
time [the time for a BH radiating at the Eddington luminosity to e-fold in
mass; we set $\epsilon/(1-\epsilon)=0.1$ below]. With the decline of material
supply, the BH growth enters into the second (or ``supply limited'') phase and
the nuclear luminosity is expected to decline below the Eddington luminosity.
Below we assume that the accretion rate in the second phase evolves as follows:
\be
\dot{\cal M}\bol(\tau)=\dot M\Edd(M_I)\left(\frac{\tau-\tau_I+\tau_D}{\tau_D}\right)^{-\gamma}, \qquad \tau\ge\tau_I,
\label{eq:Lphase2}
\ee
where $\dot{\cal M}\bol(\tau)\propto\tau^{-\gamma}$ at $\tau-\tau_I\gg\tau_D$
and $\tau_D$ is the characteristic transition timescale from the first phase to
the power-law declining stage of the accretion rate.  We assume that the two
phases appear only once for an AGN.  The final BH mass obtained after the
nuclear activity is given by
\begin{eqnarray}
M_f & = & M_I+\int_{t_I}^{\infty}(1-\epsilon)\dot{\cal M}\bol(\tau)d t \nonumber \\
& = &\left[1+\frac{\tau_D}{(\gamma-1)\tau\Sp}\right]M_I,
\label{eq:Mf}
\end{eqnarray}
where the efficiency $\epsilon$ is assumed to be the same as the efficiency in
the first phase.  Here we do not consider that the accretion process may
transit to a radiatively inefficient mode when the accretion rate becomes low
enough (i.e., when the Eddington ratio is $\la 0.001$;
\citealt{NY94,BB99}).
Such rates are in any case below the accretion rate range studied in this paper
(see Fig.~\ref{fig:distr}).  For simplicity, below we assume that the
timescales $\tau_I$ and $\tau_D$ are independent of $M_f$. 

According to the accretion rate evolution model assumed above and equations
(\ref{eq:PLM})--(\ref{eq:Phi}), $\bar{\Phi}(\dot m\OIII,\sigma)$ is roughly
constant with $\log\dot m\OIII$ if most AGNs are in the first phase and is
proportional to $\dot m\OIII^{-1/\gamma}$ if most AGNs are at the late
power-law declining stage of the accretion rate [here $P(\log M_f|\sigma)$
and $P(\log L\OIII|L\bol)$ are assumed to follow Gaussian distributions; see
also \S~\ref{sec:obs}].  If the scatter of the
distributions $P(\log M_f|\sigma)$ is sufficiently small and the bolometric
correction $f\bol$ is roughly a constant, $\dot m\OIII$ has a maximum given
by
\be
\dot m_{\max}\equiv [1+(\tau_D/\tau\Sp)/(\gamma-1)]^{-1}
\label{eq:dotmmax}
\ee
when the AGN is transiting from the first phase to the second phase.

\section{Application to observations}\label{sec:obs}

The SDSS has obtained a large sample of 33,589 narrow-line AGNs in the local
universe (http://www.mpa-garching.mpg.de/SDSS/Data/agncatalogue.html).  Details
of the sample selection and the derivation of the AGN properties used in this
paper [$L\OIII$, $\sigma$, $\mu_*$ (the stellar surface mass density)] are
given in \citet{K03a,K03b} and \citet{H04}.  In this section we fit the
accretion rate evolution models assumed in \S~\ref{sec:methods}
(eqs.~\ref{eq:Lphase1} and \ref{eq:Lphase2}) to these observed AGNs.  (Note
that Seyfert I AGNs are not included in this sample because for these objects
the AGN itself may have a significant impact at least on the estimate of the
velocity dispersion.  Possible effects on our conclusions if including Seyfert
I galaxies will be discussed at the end of this section.)

\begin{figure} \epsscale{0.5} 
\plotone{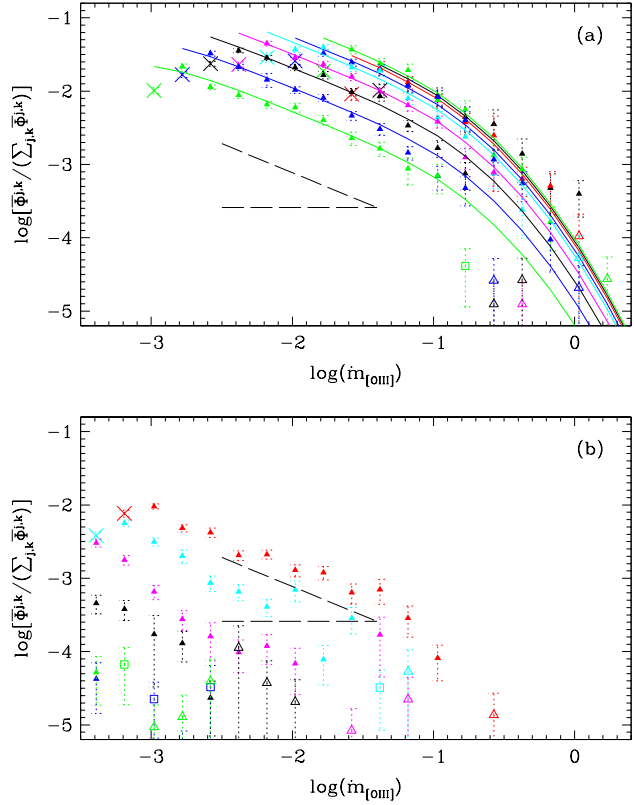}
\caption{Distribution of the normalized accretion rate $\dot m\OIII$ in
nearby AGNs ($z<0.3$). The difference between $\dot m\OIII$ and the
conventionally defined Eddington ratio is described in \S~\ref{sec:methods}.
Different
colors represent different velocity dispersion bins
$[\log\sigma_j,\log\sigma_{j+1}]$ ($j=1,2,...$) with $\sigma_1=70\kms$ and
bin size $\Delta\log\sigma=0.05\dex$.  In panel (a), from top to bottom, the
colors black, red, green, blue, cyan, magenta, black, blue, and green
correspond to $j=1,2,...,9$, respectively ($\sigma_{10}\simeq200\kms$); and in
panel (b), the colors red, cyan, magenta, black, blue, and green correspond to
$j=10,11,...15$, respectively. The sum $\Sigma_{j}$ in the normalization of the
distribution is over $j=1,2,..9$ in panel (a) and over $j=1,2,...15$ in panel
(b).  The observational distribution obtained from nearby Seyfert II galaxies are drawn
as points (triangles, crosses, or squares) at the middle point of each
$\log(\dot m\OIII)$ bin and their error bars are shown by dotted lines. For
open triangles, only one AGN is in each bin; and for open squares, only two
AGNs are in each bin. The falloff of crosses at the low-$\dot m\OIII$ end is
caused by the luminosity cut at $L\OIII=L\OIII{_{,\min}}$.  The binned model
distributions (only for $j=1,2,..9$ in panel a), obtained with
$(\tau_I/\tau\Sp,\gamma,\tau_D/\tau\Sp)=(1,1.26,3.1)$, are connected by solid
curves (for view clarity, the binned distributions are not drawn in the form of
a histogram). The observational distribution at the accretion rates $10^{-3}\la
\dot m\OIII\la 0.1$ is well consistent with a power-law distribution. 
The tilted dashed lines are the reference lines for the power-law distribution
with a slope of $-1/\gamma$.
The horizontal dashed lines are the reference lines illustrating a flat
distribution predicted by an exponential declining form of the accretion rate
evolution in the second phase of the accretion model.  See details in
\S~\ref{sec:obs}. } \label{fig:distr} \end{figure}

We consider AGNs with $L\OIII\ge L\OIII{_{,\min}}=10^6L_\odot$,
$\sigma\ge\sigma_{\min}=70\kms$, and $\mu_*>3\times 10^8\msun\kpc^{-2}$.  A
sample of fainter AGNs is incomplete because at larger distances, the fiber
samples more of the stellar light of the host galaxy and weak emission lines
become difficult to detect. The lower limit on velocity dispersion is set by
the instrumental resolution of the SDSS spectrograph.  The galaxies with lower
stellar surface mass densities are mainly disk-dominated \citep{K03a}, and
their velocity dispersions may be substantially affected by the disk
components. (Note that the quantity $L\OIII$ used here is not
extinction-corrected. The extinction effect may be implicitly included in the
scatter of the bolometric correction used in the calculations below. See more
discussions in \citealt{H04}).  We get the observational distribution
$\bar{\Phi}\kj\obs(\dot m\OIII,\sigma)$ ($j=1,2,..$; $k=1,2...$) by counting
the AGNs in such bins: (i) the bin of the velocity dispersions is
$[\log\sigma_j,\log\sigma_{j+1}]$, where
$\log(\sigma_j/\sigma_{\min})=(j-1)\Delta\log\sigma$ and the bin size
$\Delta\log\sigma=0.05\dex$.
(ii) The bin of the
normalized [OIII] line accretion rate is $[\log\dot m\OIII{_{,k}},\log\dot
m\OIII{_{,k+1}}]$, where $\log(\dot m\OIII{_{,k}}/\dot
m\OIII{_{,\min}})=(k-1)\Delta\log\dot m\OIII$, $\log\dot m\OIII{_{,\min}}=\log
L\OIII{_{,\min}}-\log L\Edd(\langle M_f\rangle)+\log f'\bol$, and the bin size
$\Delta\log\dot m\OIII=\Delta\log\sigma\times d\log[M'_f(\sigma)]/d\log\sigma
\simeq 0.20\dex$. Each observed AGN $i$ has a weighted factor $1/V_{\max,i}$,
where $V_{\max,i}$ is the volume over which the object would have been
detectable \citep{S68}.  The quantity $\bar{\Phi}\kj\obs(\dot m\OIII,\sigma)$
is obtained by summing up the weighted factors ($\sum_i 1/V_{\max,i}$) of AGNs
in each bin, and its error $\delta\bar{\Phi}\kj\obs$ is estimated by $(\sum_i
1/V_{\max,i}^2)^{1/2}$ (see points and their dotted error bars
in Fig.~\ref{fig:distr}).  Note that the realistic error may not be well
represented by the estimate above if the number of AGNs in a bin is small.
Below in the fit we do not use the points with only one (open triangles in
Fig.~\ref{fig:distr}) or two (open squares) AGNs in their bins.  In each velocity
dispersion bin, the falloff of $\bar{\Phi}\kj\obs(\dot m\OIII,\sigma)$ at the
low-accretion-rate end (see crosses in Fig.~\ref{fig:distr}) is caused by the
luminosity cut at $L\OIII=L\OIII{_{,\min}}$ and these crosses are not used in
the fit, either.

We estimate the model distribution of $\bar{\Phi}\kj\mod(\dot m\OIII,\sigma)$
by first using equation (\ref{eq:Phit1}) to obtain $\bar{\Phi}\mod(\dot
m\OIII,\sigma)$, and then binning the distribution and normalizing it by
setting $\sum_{k}\bar{\Phi}\kj\mod(\dot
m\OIII,\sigma)=\sum_{k}\bar{\Phi}\kj\obs(\dot m\OIII,\sigma)$,
where the sums are taken over the bins with $\bar{\Phi}\kj\obs(\dot
m\OIII,\sigma)\ne 0$ (but excluding the bins of crosses and open points
shown in Fig.~\ref{fig:distr}; similarly for the sum in $\chi^2$ below). We
assume that $P(\log M_f|\sigma)$ follows a Gaussian distribution with an
intrinsic scatter of $0.2\dex$ around $\langle \log M_f|\sigma\rangle$ (the
value of the scatter is set with some degree of arbitrariness, since as
mentioned in \S~\ref{sec:methods} the intrinsic scatter is difficult to
distinguish from measurement errors but is less than $0.27\dex$; \citealt{T02}.
See also discussions below).  The distribution of
$P(\log L\OIII|L\bol)$ is assumed to be independent of $L\bol$ and also follow
a Gaussian distribution with $\langle \log(1/f\bol)\rangle=\langle\log
(L\OIII/L\bol)\rangle\sim \log(1/3500)$ and a scatter of $\sim 0.38\dex$
\citep{H04}.  We use the $\chi^2$ statistic to fit the model distribution to
the observational distribution of $\bar{\Phi}\kj\obs(\dot m\OIII,\sigma)$
($j=1,2,..$; $k=1,2,...$).  The best-fit parameters are obtained by
minimizing
$\chi^2\equiv\Sigma_{j,k}[(\log\bar{\Phi}\kj\mod-\log\bar{\Phi}\kj\obs)/\delta\log\bar{\Phi}\kj\obs]^2$.
We use $P_{\chi^2}$ to denote
the probability of $\chi^2$ being higher by chance if the data were drawn from
the model.  Our a priori definition of an acceptable fit is one with
$P_{\chi^2}>0.01$.
For different $\tau_I/\tau\Sp=5,1,0.1,0$, we obtain the best fits for
parameters $(\gamma,\tau_D/\tau\Sp)$.
We find that if the observational distribution only with $\sigma_1\le\sigma\le\sigma_{10}\simeq 200\kms$ (the range of the black hole mass
$M_f'(\sigma)$ over the nine bins of the velocity dispersion is $2.0\times
10^6$--$1.3\times10^8\msun$; see Fig.~\ref{fig:distr}a) is included in the fit,
all the best fits are acceptable with $P_{\chi^2}\simeq 0.02$;
if the observational distribution only with $\sigma_1\le\sigma\le\sigma_9\simeq
176\kms$ is included in the fit, the best-fit parameters are not affected
significantly but with an improved $P_{\chi^2}\simeq 0.07-0.10$;
and if adding the observational distribution with higher
velocity dispersions $\sigma\ge\sigma_{10}$ (see Fig.~\ref{fig:distr}b)
in the fit, no fits are acceptable (since the distribution
may be complicated by some other factors, not solely by the accretion
evolution, which will be further discussed below).

\begin{figure} \epsscale{1.0} 
\plotone{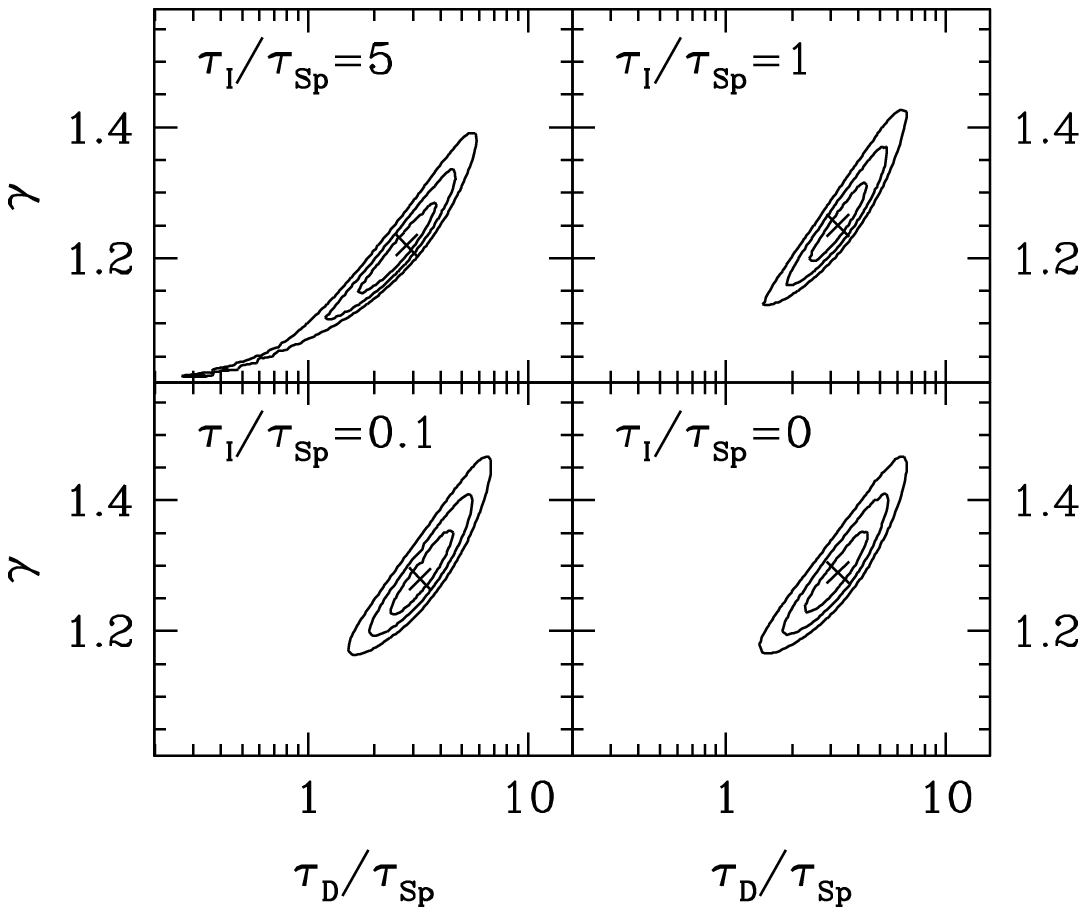}
\caption{Contours of the confidence levels for the best-fit parameters
$\gamma$ and $\tau_D$. The velocity dispersions of the AGNs used in the fit 
are in the range [$\sigma_1,\sigma_{10}$]$\simeq$[70, 200]$\kms$.
The best-fit parameters are shown by crosses.
The contours enclose the 68.3\%, 95.4\%, and 99.7\% confidence regions on them
in each panel. See details in \S~\ref{sec:obs}.
}
\label{fig:contour}
\end{figure}

For the observational distribution with $\sigma_1\le\sigma\le\sigma_{10}$, the
contour plots of confidence levels for the best-fit parameters are shown in
Figure~\ref{fig:contour}. The obtained best-fit parameters (see crosses in
Fig.~\ref{fig:contour}) are insensitive to $\tau_I$ and the average of the best
fits over four cases of $\tau_I/\tau\Sp$ is
$(\gamma,\tau_D/\tau\Sp)\simeq(1.26,3.1)$. We take the errors of the fit
parameters as $\delta\gamma\simeq\pm0.1$ and
$\delta(\tau_D/\tau\Sp)\simeq\pm1$, which roughly correspond to the 68 per cent
confidence contours (i.e., $\chi^2-\chi^2_{\min}=2.30$) shown in
Figure~\ref{fig:contour}.  We show the model distribution
$\bar{\Phi}\kj\mod(\dot m\OIII,\sigma)$ ($j=1,2,...9$) obtained with
$(\tau_I/\tau\Sp,\gamma,\tau_D/\tau\Sp)\simeq (1,1.26,3.1)$ as solid lines in
Figure~\ref{fig:distr}a.

As shown in Figure \ref{fig:distr}a, the observational accretion rate
distribution at $10^{-3}\la \dot m\OIII\la 0.1$ are well consistent with a
power-law distribution, and the parameter $\gamma$ is mainly determined by the
slope of the distribution ($-1/\gamma$). If assuming an exponentially declining
form of the accretion rate evolution in the second phase of the accretion model
instead of the power-law declining form, all fits will be unacceptable.  This
can be easily understood, since the exponentially declining form predicts a flat
accretion rate distribution (see eq.~\ref{eq:PLM}), as illustrated by the
horizontal dashed line in Figure \ref{fig:distr} (assuming no significant
cosmic evolution of the nuclear activity triggering rate).

The best-fit parameter $\tau_D$ is mainly determined by the observed turnover
of the power-law distribution of the accretion rates at the
high-rate end ($\dot m\OIII\sim 0.1$; see also $\dot m_{\max}$ in
eq.~\ref{eq:dotmmax}).  In Figure~\ref{fig:distr}, although there exist AGNs
with $\log \dot m\OIII>\log \dot m_{\max}\simeq -1.1$, the existence of these
AGNs does not necessarily mean that they are accreting at a super-Eddington
rate with $\dot M\bol>\dot M\Edd (M\bh)$ [here $\dot M\Edd(M\bh)$ is the
conventionally defined Eddington accretion rate for the BH mass $M\bh$ that is
powering the nuclear activity in the AGN, not the final mass $M_f$ after the
nuclear activity]. In the accretion evolution model used here, we always have
$\dot M\bol<\dot M\Edd (M\bh)$.  The existence of $\dot m\OIII>\dot m_{\max}$
is caused by the non-zero scatter of the $M_f-\sigma$ relation and the
bolometric correction.  The best-fit parameter $\tau_D$ may be affected by the
values of the scatter in the $M_f-\sigma$ relation and the bolometric
correction. For example, if the intrinsic scatter of $P(\log M_f|\sigma)$ is
$0.27\dex$, the best fits are $(\gamma,\tau_D/\tau\Sp)\simeq(1.24,3.5)$; if
the intrinsic scatter is zero, no fits are acceptable, but the best fits for
the observational distribution with $\sigma_1\le \sigma\le\sigma_9$ are
acceptable with $(\gamma,\tau_D/\tau\Sp)\simeq(1.23,2.1)$.  Precise
understanding of the evolution of accretion processes would require precise
measurements of this scatter (see also \citealt{YL04a}).  

As mentioned above, the observational sample used here does not include Seyfert
I galaxies. The fraction of type II AGNs in the SDSS is a weak function of AGN
luminosity, decreasing from $\sim 60\%$ to $\sim 30\%$ over the range
$L\OIII\simeq 10^{5.5}-10^{8.5}\Lsun$ \citep{H05,H04}.  Thus, the fraction of
AGNs with high accretion rates (e.g., at $\dot m\OIII\sim 10^{-1}-1$; see
Fig.~2) could be underestimated here, which will mainly affect the estimate of
$\tau_D$. After including type I galaxies, the best-fit parameter $\tau_D$
would decrease and $\gamma$ would increase slightly, resulting in an increase
in $\dot m_{\max}$ (eq.~\ref{eq:dotmmax}).  If we assume that the dependence of
the  fraction of type II AGNs on accretion rate is the same as the dependence
on  luminosity, after including type I AGNs,  we find that the distributions
(at the high accretion rates) shown in Figure 1 move upward by at most $\log
2\simeq 0.3\dex$. The value of $\tau_D/\tau\Sp$ would then have to change from
3.1 to 2.2 in order to compensate.

The best-fit $\tau_D$ above is obtained by using
the number distribution of AGNs as a function of  accretion rate.  Generally,
the timescale related to the length of the nuclear activity may also be
constrained by comparing the number density ratio of AGNs to their remnants
(e.g., Yu \& Tremaine 2002). The predicted ratio of the number density of
nearby AGNs to nearby inactive galaxies is affected by uncertainties  in the
triggering history of AGNs at high redshifts and hence does not help us obtain
a more precise constraint.

As mentioned above, if adding the observational distribution with high velocity
dispersion $\sigma\ge \sigma_{10}\simeq 200\kms$ (see Fig.~\ref{fig:distr}b) in
the fit, no fits are acceptable.  A full explanation for their detailed
distribution would need to involve at least the following four factors. (a)
These objects generally have relatively low normalized accretion rates, and
thus their accretion rate distribution could be affected by the change of the
efficiency $\epsilon$ due to the transition from the thin-disk accretion mode
to the radiative inefficient accretion mode. (b) Host halos or galaxies of
high-mass BHs generally have different major merger histories than those of
low-mass BHs.  The AGNs with high velocity dispersions generally have high BH
masses, low normalized accretion rates, and probably long ages of the nuclear
activities, and thus their distribution could be more likely to be affected by
the cosmic evolution of the nuclear activity triggering history.  (c) The low
accretion rate in these objects (e.g., $\dot{m}\la 10^{-3}$) may be simply
caused by Bondi accretion. Note that the observational accretion rate
distribution at the low accretion rate end shown in Figure \ref{fig:distr}b has
indicated a steeper increase than the slope $-1/\gamma$. (d) The assumption
that $\tau_D$ are independent of $M_f$ may be violated for a wide range of
$M_f$.

\section{Comparison with accretion disk evolution models}\label{sec:models}

The process that removes the angular momentum of gas in a galaxy and drives it
into the very central region is still not well understood.  It  may require a
hierarchy of mechanisms, handing off to one another as the material moves
inward. Here we consider the ``last stop'' of the ``handing-off''. We
hypothesize that some mechanism has moved gas to the central parsec in the
galaxy, where it will then diffuse to the central BH through viscous transport
(or specifically by the magneto-rotational instability; e.g., \citealt{BH98})
in an accretion disk.
The surface mass density, $\Sigma$, of an accretion disk with a central,
dominant gravitating point mass and with viscosity $\nu$ evolves as a function
of time $\tau$ and radius $R$ through the equation
\be
\frac{\partial\Sigma}{\partial \tau}=\frac{3}{R}\frac{\partial}{\partial R}
\left[R^{1/2}\frac{\partial}{\partial R}(\nu\Sigma R^{1/2})\right].
\ee
If the viscosity is of the form $\nu(R,\Sigma)\propto R^m\Sigma^{n}$ (where $m$ and
$n$ are constants), this equation has solutions of self-similar form
\citep{P81} as follows:
\be
\frac{\Sigma(R,\tau)}{\Sigma_0}=\left(\frac{\tau}{\tau_0}\right)^\eta 
f\left[\left(\frac{R}{R_0}\right)\left(\frac{\tau}{\tau_0}\right)^\xi\right],
\label{eq:Sigma}
\ee
where the values of $\eta$ and $\xi$ are determined by the exponents $m$ and
$n$ in the function  $\nu(R,\Sigma)$ and also by the  detailed boundary
conditions of the accretion disk.  $\Sigma_0$, $R_0$, and $\tau_0$ are scale
variables with  arbitrary  dimensions and they                 satisfy
$\tau_0=R_0^2/\nu(R_0,\Sigma_0)$.  The self-similar evolution of accretion
disks has also been calculated using  numerical simulations (e.g.,
\citealt{LP87,CLG90,P91}).  For an accretion disk with negligible torque at
both the origin $R=0$ and its outer boundary (so that its total angular
momentum is conserved during its evolution), the rate at which the mass of the
disk decreases is given by a    power law of the  form $\dot
M_d(\tau)=\frac{d}{d\tau} (\int 2\pi\Sigma R
dR)\propto\tau^{-\frac{38+18a+4b}{32+17a+2b}}$, where $a$ and $b$ are the
exponents in the opacity law $\kappa(\rho,T)=\kappa_0\rho^{a}T^{-b}$ (e.g.,
\citealt{CLG90}).  If $a=b=0$ (Thomson opacity), we have $\eta=-15/16$,
$\xi=-3/8$, $f(x)=(28)^{-3/2}x^{-3/5}(1-x^{7/5})^{3/2}$ in equation
(\ref{eq:Sigma}), and the mass accretion rate evolves as
\be
|\dot M_d(\tau)|=\frac{(16/3)M_{d,0}}{\tau_0} \left(\frac{\tau}{\tau_0}\right)^{-19/16},
\ee
where $M_{d,0}=(28)^{-3/2}\frac{4\pi}{7}R_0^2\Sigma_0$,
and the diffusion timescale of the disk
\be
\tau_0=\frac{R_0^2}{\nu(R_0,\Sigma_0)}=(0.1-1.6)\times 10^8\yr
\left(\frac{R_0}{0.3-1\pc}\right)^{7/3}\left(\frac{M\bh}{10^7\msun}\right)^{1/3}
\nonumber \\ \times\left(\frac{\alpha}{0.1}\right)^{-4/3}\left(\frac{M_{d,0}}{10^7\msun}\right)^{-2/3},
\label{eq:tau0}
\ee
where $\alpha$ is the commonly used parameter related to the viscosity in the
standard accretion disk model, and $R_0$ may be taken as the size of the
accretion disk or the location of its reservoir. The value of $\tau_0$ is sensitive
to $R_0$ and the reference value of $R_0$
in equation (\ref{eq:tau0}) is roughly consistent with the
observations of the accretion disks in several nearby AGNs (e.g., NGC4258,
\citealt{M95}; NGC1068, \citealt{LB03}; and NGC 3079, \citealt{KGM05}).
 If $a=1$ and $b=7/2$ (Kramers opacity), we have $|\dot
M_d|\propto\tau^{-5/4}$. 

Note that the best-fit parameter $\gamma=1.26\pm0.1$ for AGNs with
$\sigma\simeq70-200\kms$ is consistent with the exponent (-1.18 or -1.25) of
$\tau$ in $\dot M_d$ expected from the self-similar solutions above.
The diffusion timescale $\tau_0$ in equation (\ref{eq:tau0}) is also roughly
consistent with $\tau_D$ obtained from observations. These consistencies between the
observations and the theoretical expectations above are an encouraging result.
They support the hypothesis that the observed diversity in the Eddington ratios
of AGNs results from an evolutionary sequence of accretion processes. They also
suggest that the accretion disks of most Seyfert galaxies (with
$\sigma\simeq70-200\kms$ and $\dot m\ga 10^{-3}$) are currently in the
self-similar phase or are transiting toward this phase.  It is interesting to
note that the self-similar evolution of disk evolution is also supported by
observations of accretion disks around T Tauri stars \citep{H98}.

\section{Discussions}\label{sec:dis}

Despite the encouraging consistency above, the self-similar solution of the
accretion disk evolution shown in \S~\ref{sec:models} is based on highly
simplified accretion disk models. For example, the effects on the disk
evolution by disk winds or by infalling material that is continuously deposited
onto the disk are not considered.  The role of possible disk instabilities is
not considered, either. More complete investigations on the long-term evolution
of accretion disks are needed.

There are also further issues to be considered:

\begin{itemize}

\item {\em Self-gravitating disks and star formation:} With the best-fit
parameters obtained for the accretion evolution in \S~\ref{sec:obs}, the disk
is self-gravitating in its early evolutionary stage and the central BH mass
becomes dominant in its  late evolutionary stage. We note that this
evolutionary pattern of the disk mass is probably compatible with current
observations of the accretion disks in a few AGNs. For example, NGC3079
\citep{KGM05} and NGC1068 \citep{LB03} have Eddington ratios close to $1$,
which may correspond to the first ``demand-limited'' phase or a transition to
the self-similar evolution of the second phase. Their accretion disks appear to
be thick with masses larger than or comparable to their central BH masses.
NGC4258 has an Eddington ratio $\la 0.01$ (\citealt{GNB99} and references therein), so it may be in the
late stage of self-similar evolution, and its accretion disk is thin with a
mass that is  much smaller than its BH mass \citep{M95}. The evolution of a
self-gravitating disk may not exactly follow the simple prescription adopted in
this paper. Recently, \citet{FBD04} simulated the evolution of self-gravitating
tori around a massive BH and found that they quickly evolve to a dual structure
with an inner thin Keplerian disk fed by an outer thicker self-gravitating
disk (see also \citealt{BL99,LR04,LR05}).  Further investigation of the long term evolution of self-gravitating
disks is crucial for  understanding the transition to the self-similar phase.
In addition, in a self-gravitating disk, star formation may be important
because of gravitational instabilities that develop in the disk \citep{G03,L03}.
However, the growth of the central BH will  consume a substantial fraction of
the disk mass.  The fraction of the disk mass that goes into stars rather than
the BH is worthy of further investigation.

\item {\em The $M_f-\sigma$ relation in normal galaxies and the $M\bh-\sigma$
relation in AGNs:} As mentioned in \S~\ref{sec:methods}, in our model the
$M_f-\sigma$ relation in normal galaxies is used as a boundary condition for BH
mass growth in AGNs, and BH masses in AGNs are lower than the expectation from
the $M_f-\sigma$ relation. In observations some AGNs with sufficiently low
Eddington ratios may be included in normal galaxies.  As a self-consistency
check for our model, here we show that the best-fit parameters for BH growth
obtained in \S~\ref{sec:obs} do not contradict the tightness of the
$M_f-\sigma$ relation. For example, using equations (\ref{eq:Lphase2}) and
(\ref{eq:Mf}), for $(\gamma, \tau_D)=(1.26,3.1)$, when the Eddington ratio of
an AGN declines to below $10^{-3}$, the logarithm of the ratio of its
central BH mass to its final mass is $-0.18\dex\la\log(M\bh/M_f)<0$, which
does not exceed the scatter of the $M_f-\sigma$ relation used in the model.  In
addition, considering that most of the supplying material may be blown away as
outflows when a BH accretes material via the ADIOS (adiabatic inflow-outflow
solutions) mode at low accretion rates \citep{BB99}, the difference between the
BH masses in AGNs with low Eddington ratios and their final masses will be
smaller.

For the best fits $(\gamma, \tau_D)=(1.26,3.1)$, BH masses in most AGNs
($\sigma \simeq 70-200\kms$) with Eddington ratio in the range 1--0.1 are about
0.08--0.3 times their final mass $M_f$. (Note that here at a given velocity
dispersion and luminosity, most AGNs are in the second evolutionary stage,
rather than in the first stage in which the nuclear luminosity exponentially
increases.) However, by studying 15 dwarf Seyfert I galaxies
($\sigma\simeq30-80\kms$), \citet{BGH05} found that the BH masses in these
AGNs, $M\bh$ (estimated from the empirical relations among the BH mass, the
nonstellar $5100\AA$ continuum luminosity, the broad-line region radius, and
the broad-line width in \citealt{K00}), are consistent with the $M_f-\sigma$
relation and these AGNs have Eddington ratios in the range 0.1--1.
Currently it is hard to answer whether the evolution model in our paper
contradicts the observations or not, because the answer should take into
account at least the following uncertainties.  (i) Usually, BH masses $M\bh$
used to derive the empirical relations in AGNs are estimated through the
reverberation mapping technique with some assumptions on the kinematics of the
broad-line emission clouds \citep[e.g.,][]{W99,K00}. The reverberation mapping
technique could either systematically overestimate or underestimate the BH mass
by a factor of 3 or so \citep{K01}.  (ii) It is possible that the sample in
\citet{BGH05} biases toward detection of AGNs with the highest mass black holes
for a given $\sigma$.  (iii) The constraint on $\tau_D$ obtained in
\S~\ref{sec:obs} may not be strict (e.g., because of the exclusion of Seyfert I 
galaxies).  If $\tau_D$ is smaller (e.g., $\sim 1\tau\Sp$), the BH masses in
most AGNs with Eddington ratios in the range 1--0.1 will be closer to their
final mass (e.g., $\sim0.2M_f- 0.4M_f$).  (iv) The dwarf galaxies in \cite{BGH05}
are in a velocity dispersion range different from the SDSS AGNs and may not
follow the same BH growth parameters.

In addition, we note that the BH masses in AGNs measured through the
reverberation mapping technique by \citet{O04} are calibrated to follow the
$M_f-\sigma$ relation in normal galaxies by multiplying them by a factor of
1.8.  \citet{O04} argue that the calibration factor could be caused by a
nonisotropic velocity distribution of the broad-line emission clouds; however
an alternative explanation could be that there exists an offset between the
$M\bh-\sigma$ relation in AGNs and the $M_f-\sigma$ relation in normal
galaxies, which just represents BH mass growth during the nuclear activity
phases \citep{YL04b}. It is worthwhile to study whether such an offset really
exists, which would provide strong constraints on BH growth models and the
physical origin of the $M_f-\sigma$ relation. 

\item {\em Binary BHs (BBHs) and coevolution of galaxies and QSOs/AGNs:} In the
current coevolution model of galaxies and QSOs \citep{KH00}, the nuclear
activity of galaxies are triggered by major mergers of galaxies.  Mergers of
galaxies may form a BBH if there exists a massive BH in each of the merging
galaxies \citep{BBR80,Y02,VHM03}. The evolution of the accretion disk around a
massive BBH may be different from that around one BH (e.g., because of
different boundary conditions of the disk) and depend on the orientation of the BBH orbital
plane and BH spins. For example, studies of the time-dependent evolution of
the accretion disk around the primary of a binary star with the outer disk
radius tidally truncated by the secondary also yield  self-similar solutions
but with a steeper time dependence of the accretion rate $\dot M_d\propto
\tau^{-2.5}$ (Thomson opacity) or $\tau^{-3.3}$ (Kramers opacity)
(\citealt{LS00}; see also a different self-similar solution for an external
disk around a binary star by \citealt{P91}).  The consistency of the slope of
the accretion rate function with the expectation of our simple model
may suggest that galaxy mergers have not been sufficiently  efficient  in faint
AGNs to form a large population of  BBHs, or that galaxy triaxiality or gas
dynamics has caused BBHs to merge very quickly.

The AGN luminosity function, defined by the comoving number density of AGNs per
unit luminosity at a given redshift, is proportional to $\int {\dot
m}^{-1}\Phi(\dot m,\sigma)d\log\sigma$ with $\dot m\propto
L/L\Edd[M_f'(\sigma)]$.  The luminosity function at the faint end, if fitted by
a power-law form $\propto L^\beta$, is expected to have the slope $\beta\la
-1-1/\gamma$ (i.e., $\beta\la -1.85,-1.80,-1.40,-1.30$ for
$\gamma=1.18,1.25,2.5,3.3$, respectively), where the symbol $\la $ is
expected instead of the symbol $\simeq$ because the break of the power-law accretion
rate distributions ($\dot m\simeq \dot m_{\max}$) at the high-rate end
corresponds to different break luminosities for AGNs with different velocity
dispersions, and thus the faint end of the luminosity function consists of not only
AGNs with accretion rates lower than the break rate but also some with rates
higher than the break rate, which may steepen the slope.  According to the fit
for the accretion rate distribution of nearby AGNs, the expectation for their
luminosity function at the faint end, $\beta\la -1.8$, is consistent with the
fit for the observational luminosity function by \citet{H05}.  Observations of
the QSO luminosity function indicate a relatively flat slope
$\beta\simeq-1.45\pm0.03$ at the faint end \citep{R05}, which might suggest the
existence of BBHs in a significant fraction of QSOs since $-1.8<\beta<-1.3$.
It is worthwhile incorporating the evolution of BBHs and the evolution of
accretion disks  in models for  the coevolution  of galaxies and QSOs/AGNs.
One could then make joint predictions for the evolution of the QSO/AGN
luminosity function as well as their accretion rate distribution function for
comparison with observations.

\item {\em Possible difference between AGNs and bright QSOs:} The luminosity
evolution of bright QSOs ($L\bol\ga 10^{46}\ergs$) is constrained from
observations through a relation between the BH mass function in nearby normal
galaxies and the time integral of the QSO luminosity function in \citet{YL04a}.
The characteristic timescale for the luminosity of the QSO to decline during
its second evolution phase was shown to be very short ($<0.3\tau\Sp$;
\citealt{YL04a}), significantly shorter than the timescale obtained for  AGNs
in this paper.\footnote{The shortness of the characteristic timescale for QSOs
can be explained as follows : (i) for a given  BH mass, the peak nuclear
luminosity decreases with increasing  characteristic timescale (see
eq.~\ref{eq:Mf}); (ii) because the local BH mass function decreases
exponentially at the high-mass end, a long characteristic timescale results in
QSO number densities that are much lower than the observed ones.  Note that the
characteristic timescale here is different from the ``lifetime'' of the nuclear
activity phase. The value of the QSO lifetime is sensitive to the detailed
definition of  ``nuclear activity''(e.g., the lower threshold of the nuclear
luminosity for a galaxy to be classified as ``active''). The  determination of
lifetime is  more likely to be affected by uncertainties of the QSO luminosity
function at the faint end. See more discussions in \citet{YL04a}.} This
difference between AGNs and bright QSOs is slightly smaller if we include type
I AGNs, but it still persists (see discussion in \S~\ref{sec:obs}).
In \citet{YL04a}, the luminosity in the second phase is assumed to decrease
exponentially and not according to  a power law.  If the QSO luminosity
evolution is assumed to have the same evolutionary form as described in
equations (\ref{eq:Lphase1}) and (\ref{eq:Lphase2}), an even  smaller $\tau_D$
(and/or larger $\gamma$) would be required.  According to equation
(\ref{eq:tau0}), $\tau_0$ (or $\tau_D$) is highly sensitive to the location of
reservoir of accreting materials ($\propto R_0^{7/3}$) but depends weakly on
$M\bh$ and $M_d$. A smaller $\tau_D$ for QSOs implies that the disk size or the
location of the material that fuels the    BH in QSOs is not larger than in
AGNs; this is surprising because the  BHs in QSOs are one or two orders of
magnitude larger than the BHs of the AGNs that we have studied in this paper.
This may provide interesting constraints on the fueling mechanisms that remove
the angular momentum of materials at galactic scales and cause  them to
circularize at the parsec scale.  A larger $\gamma$ might suggest the existence
of BBHs in QSOs (see also the discussion above).

\item {\em Unified Models of AGNs: adding the effect of an evolving accretion
disk } In current unified models of AGNs, orientation effects are proposed to
explain the diversity of AGNs/QSOs \citep{A93}.  The accretion rate and the
Eddington ratio have also been proposed as important parameters for
understanding why AGNs are so diverse \citep{B02}.  In this paper, we connect
the diversity of the Eddington ratio in AGNs with the evolution of accretion
disks, and we suggest that different kinds of AGNs can be partly unified by
considering the time sequence of the disk evolution.

\begin{itemize}

\item{\em Lack of a torus in very weak AGNs:} The current AGN unification model
requires the existence of a torus in an AGN, but observations show that it does
not exist in M87 \citep{WA04}. Here we propose that the lack of a torus may be
explained as an  evolutionary consequence of the accretion process.  At an
early stage, the outer disk may be thick and clumpy and play the role of a torus.
At very late stage, most of the outer disk (or the torus) is very likely to
have diffused away and the emission from the rest  of the  torus may be too
weak to detect.  M87 has a low Eddington ratio $\la 10^{-3}$ \citep{D03} and
may correspond to the very late stage of the disk evolution in our scenario.
Note that this diffusing torus scenario  is different from the receding torus
model in which luminous AGNs may have weak tori due to sublimation
\citep{L91}.  Generally, the evolution of a torus (or outer disk) will be
driven by both diffusion and sublimation. 

\item{\em Radiatively inefficient accretion:} With the decrease of the
accretion rate in the second evolutionary phase (but before it decreases down
to the Bondi accretion rate), the accretion may transit to a radiatively
inefficient mode (ADAF or ADIOS; \citealt{NY94,BB99}).  The existence of some
AGNs whose spectra need to be explained through radiatively inefficient
accretion modes (but with an accretion rate larger than the Bondi accretion
rate) may thus  also be explained as a consequence of the evolutionary sequence
described in this paper. The distribution of the normalized accretion rates  of
nearby low-luminosity AGNs could provide constraints on the transition from
the radiatively efficient accretion mode to the radiatively in-efficient mode (see
also \citealt{BC04}).

\item {\em Radio-loud vs.\ radio-quiet:} AGNs/QSOs can be classified into two
groups: radio-loud and radio-quiet. It has been a long-standing question
whether all AGNs pass through a radio-loud phase or whether only certain
objects  develop jets and become radio-loud.  \citet{B02} shows that most
radio-quiet QSOs have high Eddington ratios, while most radio-loud QSOs have
low ratios. Since the accretion rate decreases with time as the accretion disk
evolves, we propose that from the evolutionary point of view, a QSO/AGN appears
radio-quiet at its early evolution stage and appears radio-loud at its
late stage (of course, the detailed appearance is also complicated by
orientation effects and possibly by BH spins).

\end{itemize}

\end{itemize}

\section{Conclusions}\label{sec:con}

In this paper we have connected the accretion rate distribution of nearby SDSS
AGNs ($z<0.3$) with the accretion physics around massive BHs.  The observed
distribution of the AGNs (with host galaxy velocity dispersion
$\sigma\simeq70-200\kms$) follows a power-law distribution at low accretion
rates ($\dot m\OIII\simeq 10^{-3}-0.1$), which is consistent with a scenario in
which  their mass accretion rates are declining with time in a power-law form
($\dot {\cal M}\propto \tau^{-\gamma}$ with $\gamma=1.26\pm0.1$) and the accretion
process follows a self-similar evolutionary pattern, as simple theoretical
models predict. The transition timescale of the accretion to the self-similar
phase (estimated from the turnover of the observational power-law distribution
at high accretion rates) is roughly consistent with the diffusion timescale of
an accretion disk. The results suggest that the observed diversity in the
Eddington ratios of AGNs ($\sim 10^{-3}-1$) is a natural result of the
long-term evolution of accretion disks.  Some other issues deserving of further
investigation are discussed, such as a full explanation for the accretion
rate distribution of AGNs with higher velocity dispersions
($\sigma\ga200\kms$), the long-term evolution of accretion disks (including
self-gravitating disks), the BH mass versus velocity dispersion relation in
AGNs, the evolution of BBHs in QSOs/AGNs, the coevolution of galaxies and
QSOs/AGNs, and the unification picture of AGNs.

We thank the referee for thoughtful comments.
QY acknowledges support provided by NASA through Hubble Fellowship grant
\#HF-01169.01-A awarded by the Space Telescope Science Institute, which is
operated by the Association of Universities for Research in Astronomy, Inc.,
for NASA, under contract NAS 5-26555.

\end{document}